# Convolutionally Coded SNR-Adaptive Transmission for Low-Latency Communications

Mehmet Cagri Ilter  and Halim Yanikomeroglu

*Abstract*—Fifth generation new radio aims to facilitate new use cases in wireless communications. Some of these new use cases have highly demanding latency requirements; many of the powerful forward error correction codes deployed in current systems, such as the turbo and low-density parity-check codes, do not perform well when the low-latency requirement does not allow iterative decoding. As such, there is a rejuvenated interest in noniterative/one-shot decoding algorithms. Motivated by this, we propose a signal-to-noise ratio-adaptive convolutionally coded system with optimized constellations designed specifically for a particular set of convolutional code parameters. Numerical results show that significant performance improvements in terms of bit-error-rate and spectral efficiency can be obtained compared to the traditional adaptive modulation and coding systems in low-latency communications.

## I. INTRODUCTION

Minimizing error rates has been principal goal in wireless system design for decades. Infinite-length code blocks are commonly assumed in capacity calculation for any given transmission system and the invention of the capacity-approaching/achieving codes such as; turbo, LDPC and polar codes have been among the key developments in this field. However, NR has been introduced the low-latency use cases where different packet sizes, code rates and the robustness in performance against various fading effects are desired characteristics along with acceptable encoder/decoder complexity [1]. For instance, spatially coupled LDPC code has been developed for low-latency communications [2]; they are performed well at short-to-moderate packet sizes.

Constellation shapings is one of the popular radio access network (RAN) solution for capacity maximization which are based on either using non-uniformly located symbols (geometrical shaping) or non-equally probable symbols (probabilistic shaping) [3]. In the absence of optimization as a tool, most earlier constellation shaping works have been confined to cases in which the possible signal point locations were located on regular lattices [4] or which had other restrictive constraints. With the utilization of the optimization techniques, constellation shaping has increasingly been studied through the use of optimization techniques [5]–[7], and non-uniformly spaced constellations dedicated

Manuscript received November 30, 2017; revised April 1, 2018 and May 4, 2018; accepted May 26, 2018. Date of publication June 4, 2018; date of current version September 17, 2018. This work was supported in part by the Huawei Canada Co., Ltd., and in part by Natural Sciences and Engineering Research Council of Canada's (NSERC) Strategic Partnership Grants program. The review of this paper was coordinated by Prof. Y. Li. *(Corresponding author: Mehmet Cagri Ilter.)*

The authors are with the Department of Systems and Computer Engineering, Carleton University, Ottawa, ON K1S 5B6, Canada (e-mail: ilterm @sce.carleton.ca; halim@sce.carleton.ca).

Color versions of one or more of the figures in this paper are available online at http://ieeexplore.ieee.org.

Digital Object Identifier 10.1109/TVT.2018.2844019

to different transmission models emerge in the wireless standards without any regular grid constraint. Interestingly, it was also proven that the usage of uniformly spaced constellations can cause suboptimal performance in HSPA, IEEE.802.11.$a/g/n$, DVB-T2 and DVB-S2 standards [8]. It is also shown that the capacity loss resulting from the conventional uniform quadrature amplitude modulation (QAM) constellations increases with larger SNR values over bit-interleaved coded modulation (BICM) cases [9]. The use of non-uniform constellations also emerges in video broadcasting standard, ATSC 3.0, which might be the first major broadcasting standard where non-uniform constellations are deployed [10], and has been proposed in 5G NR framework studies [11], [12].

Considering low-latency use cases with existing delay constraint and shorter block sizes in packet communications, a simple design along with robustness of various fading conditions appears as key objective starting from the rise of NR concept [13] where previously used capacity-approaching codes in the standards cannot tackle desired latency requirements due to iterative decoding process [14], [15]. For instance, the advantage of convolutional encoders over the capacity-approaching codes under strict delay constraint was presented in [16]. Furthermore, it was interestingly shown that even polar codes have became less preferable than convolutional encoders in particular packet size and system complexity requirements [17]. At this point, we aim to propose a framework which makes convolutional encoder more attractive for the use in mentioned low-latency use cases. Specifically, proposed convolutionally coded model offers the use of various SNR-adaptive optimized constellations which strengthens low-delay advantage of one-shot convolutional decoding with performance gain obtained from channel and encoder-based design over Nakagami-$m$ channel.

The remainder of the paper is organized as follows: In Section II, a detailed description of the process for finding SNR-adaptive optimized irregular constellations is given. Then, Section III introduces SNR-adaptive convolutionally coded transmission model where optimized constellations for each SNR value are used and a new approach for constructing modulation and coding schemes along with those SNR-adaptive constellations is introduced. Section IV presents the performance enhancements in terms of spectral efficiency and decoding latency obtained from the proposed framework. Section V concludes on the findings.

## II. IRREGULAR OPTIMIZED CONSTELLATIONS

Error performance analysis of convolutional encoder is mainly based on the generating function which is calculated from the state transition diagram of the encoder. In general, the all-zero sequences are assumed to be transmitted in the generating function calculation over quasi-regular (QR) cases; however, many systems can be found to be irregular, especially when the encoders are paired with non-uniformly spaced constellations [18]. Motivated by the fact that being QR is not associated with either better or worse error performance, we first summarize the error performance calculation for the convolutional coder which enables search for optimized symbol locations without any predefined assumption on their locations over integer-valued Nakagami-$m$ fading channels.[1]

### A. Error Bound Over Irregular Constellation Cases

Initial step for the calculation of upper bound on bit error rate (BER), $P_b$, upper bound, via the product-state matrix technique is constructing the product-state matrix, $\mathbf{S}$ for a given encoder. Its particular entry of $\mathbf{S}$, $\mathbf{S}_{(u,v),(\bar{u},\bar{v})}$, can be expressed as

$$\mathbf{S}_{(u,v),(\bar{u},\bar{v})} = \Pr(u \to \bar{u}|u) \sum_n p_n \mathcal{I}^{\mathcal{W}(u \to \bar{u}) \oplus \mathcal{W}(v \to \bar{v})} D_{(u,v),(\bar{u},\bar{v})},$$

(1)

where the summation in (1) is over all the possible $n$ parallel transitions and $\mathcal{I}$ is a dummy variable which disappears before the derived upper bound expression. $\Pr(u \to \bar{u}|u)$ is the conditional probability of a transition from state $u$ to state $\bar{u}$ given state $u$, and $\mathcal{W}(i \to j)$ denotes the Hamming weight of the information sequence for the transition from $i$ to $j$, where $i \in \{u, v\}$ and $j \in \{\bar{u}, \bar{v}\}$ [20]. Herein, $D_{(u,v),(\bar{u},\bar{v})}$ can be interpreted as the Chernoff bound for the probability of decoding the erroneous transition $v \to \bar{v}$ rather than the correct one $u \to \bar{u}$ under maximum likelihood (ML) decoding [21], and it can be formulated as [18]

$$D_{(u,v),(\bar{u},\bar{v})} = (1 + \Omega|s - \hat{s}|^2/(4N_0 m))^{-m},$$

(2)

with $s, \hat{s} \in \chi(m, \bar{\gamma})$ for integer-valued Nakagami-$m$ fading scenarios and $N_0$ is noise variance. Herein, $m$ is the fading parameter, $\Omega$ is the average fading power, and $\chi(m, \bar{\gamma})$ denotes the constellation used for given $m$ and the average received SNR value, $\bar{\gamma}$. After obtaining each entry of $\mathbf{S}$ based on (1) and suitable ordering the product-states based on their classifications [19], $\mathbf{S}$ can be written in the form of [21]

$$\mathbf{S} = \begin{bmatrix} \mathbf{S}_{\mathcal{G}\mathcal{G}} & \mathbf{S}_{\mathcal{G}\mathcal{B}} \\ \mathbf{S}_{\mathcal{B}\mathcal{G}} & \mathbf{S}_{\mathcal{B}\mathcal{B}} \end{bmatrix},$$

(3)

then, the generating function, $T(D, \mathcal{I})$, can be formulated as [21]

$$P_b \le \frac{1}{l} \frac{\partial}{\partial \mathcal{I}} \mathbf{1}^T \mathbf{S}_{\mathcal{G}\mathcal{G}} \mathbf{1} + (\mathbf{1}^T \mathbf{S}_{\mathcal{G}\mathcal{B}})^T [\mathbf{I} - \mathbf{S}_{\mathcal{B}\mathcal{B}}]^{-1} \mathbf{S}_{\mathcal{B}\mathcal{G}} \mathbf{1} \Big|_{\mathcal{I}=1},$$

(4)

where $\mathbf{1}$ and $\mathbf{I}$ denote the unity and identity matrices, respectively.

### B. Particle Swarm Optimization

In order to determine symbol points locations which give the optimal performance for a given set of system parameters $(m, \bar{\gamma})$, the optimizer uses an metaheuristic evolutionary technique which is based on swarm intelligence. Particle swarm optimization (PSO) technique is chosen in this study because of its less computational load and fewer tuning parameters as compared to other evolutionary algorithms. The PSO already preferred in implementing the optimization framework in wireless systems, for instance the radio planning algorithm in the long term evolution (LTE) systems [22].

The PSO optimizer requires the following parameters before going through the constellation search: the fading parameter $m$, the average fading power $\Omega$, the modulation order $M$, the swarm size $P$, and optimizer parameters $(r_1, r_2, w, N_{\text{iter}})$. After initializing the positions of $P$ particles', $\mathbf{x}_i^0$, and their velocities, $\mathbf{v}_i^0$, the fitness value of each particle is calculated from (4). Then, the best value of the swarm, $g_p^0$, which is the one giving the minimum value of $P_b$, is calculated. The velocities and positions of the particles are updated as follows:

$$x_i^n = x_i^{n-1} + v_i^n$$

(5a)

$$v_i^n = v_i^{n-1} + r_1 c_1 \left(g_i^{n-1} - x_i^{n-1}\right) + r_2 c_2 \left(g_p^{n-1} - x_i^{n-1}\right).$$

(5b)

Calculations of (4) are carried out by considering these updated values and the ones giving the lowest $P_b$ values are kept as the updated particles. At the end of $N$th iteration, the $x_i^{N_{iter}}$ yields to the optimized irregular constellation, $\chi(m, \bar{\gamma})$. The used parameters for constellation search can be found in [23]. Throughout the search for symbol point locations, the following constraint on the symbol point locations is only taken into account, which guarantees the average transmitted symbol energy cannot exceed the average energy constraint, $E_s$; it is given by,

$$\frac{1}{M} \sum_{i=0}^{M-1} |s_i|^2 < E_s, \ s_i \in \mathbb{C}, \forall i.$$

(6)

Since the optimized constellations $\chi(m, \bar{\gamma})$ vary depending on $(\bar{\gamma}, m)$ and the encoder properties, each constellation need to be stored along with a label of corresponding $\bar{\gamma}$ and $m$ values for future use in the transmission. By this way, an appropriate $\chi(m, \bar{\gamma})$ can be used in the transmitter based on the channel information from the receiver when transmission occurs in a given convolutionally coded scenario.

### C. SNR-Adaptive Irregular Constellations

Before the mathematical proof presented in [24], the regular simplex constellations were considered as the unique constellation choice which maximize the minimum distance for a given constellation [25] for uncoded scenarios. Then, it was proven that one constellation might not be optimal for all SNR values. With the inspiration from uncoded scenarios, the examples of deploying multiple constellations which vary based on received statistics can be seen recent communication standards such as; ATSC 3.0 digital video broadcasting standard [10] and fiber-optical communications [7].

Since the proposed model aims the robustness against varying channel conditions, the optimizer requires to perform its search for optimized constellations, $\chi(m, \bar{\gamma})$ for a given $m$ parameter and chosen $\bar{\gamma}$ values in a given interval $\bar{\gamma} \in [\bar{\gamma}_i, \bar{\gamma}_f]$ along with implicitly considering encoder type and puncturing pattern if it exists. Meanwhile, a set of optimized constellations, $\chi(m, \bar{\gamma})$ for different values of $(\bar{\gamma}, m)$ is kept in look-up tables for future-use as illustrated in Fig. 1.

## III. SYSTEM MODEL

### A. SNR-Adaptive Convolutionally Coded Transmission

Basically, the SNR-adaptive convolutionally coded transmission model is given in Fig. 1. The information bits belonged by $l$th frame $\mathbf{b}_l = [b_{l,1} \cdots b_{l,N_b}]$ are encoded (one frame has $N_b$ information bits) and the encoded bits, $\mathbf{c}_l = [c_{l,1} \cdots c_{l,N_c}]$, are fed to the bit-to-symbol mapper in which transmitting symbols with a length of $N_s$, $\mathbf{s}_l = [s_{l,1}, \cdots, s_{l,N_s}]$, are assigned from $\chi(m, \bar{\gamma})$. The fading coefficients of the channel for $l$th frame, $\mathbf{h}_l = [h_{l,1}, \cdots, h_{l,N_s}]$, is modeled by frequency non-selective Nakagami-$m$ fast fading model with a shaping parameter $m$ and an average fading power $\Omega$. Then, the received signal for the $i$th symbol in the $l$th frame can be written as

$$r_{l,i} = h_{l,i} s_{l,i} + n_{l,i}, \qquad (7)$$

where $n_{l,i}$ is the additive white Gaussian noise (AWGN) sample with zero-mean and $N_0/2$ noise variance per dimension, $s_{l,i} \in \chi(m, \overline{\gamma})$ where the average received SNR can be explicitly defined as $\overline{\gamma} = \Omega E_s / N_0$. Here, $E_s$ denotes the average symbol energy of $\chi(m, \overline{\gamma})$ and $\Omega$ can interpreted as the path-loss term. Note that we assume that the channel coefficient stays constant during one symbol transmission and that each symbol is exposed to a different fading coefficient. In the receiver side, soft-decision Viterbi decoding is used by assuming that the perfect channel state information (CSI) and corresponding constellation, $\chi(m, \overline{\gamma})$, are known throughout the decoding process.

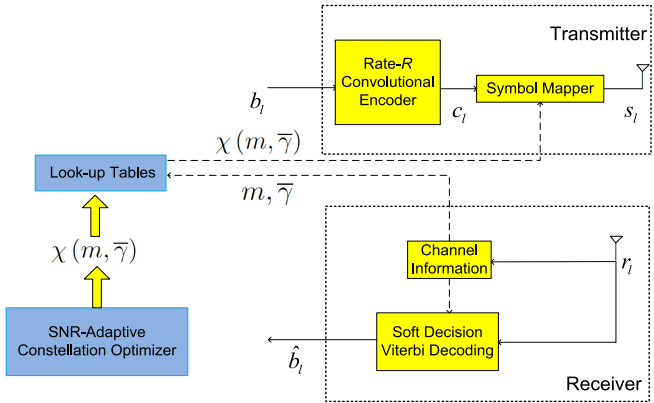

Fig. 1.   Convolutionally coded SNR-adaptive transmission model.

### B.  New Approach for Constructing MCSs

Most existing modulation and coding schemes (MCSs) are based on deploying a limited number of constellations compared to total number of available MCSs. For instance, there are 15 different modulation order and coding rate pairs where only 3 different constellations, which are $M$-QAM constellations for $M = \{4, 16, 64\}$, are used in LTE systems and the one yielding the best throughput performance under the certain block error rate threshold is selected during the transmission [26]. While increasing available MCS options brings considerable cost in computing and storage ability, higher number of MCSs can be seen in more advanced systems, as in transmission protocols with hybrid automated request (HARQ) where 27 MCS options are available [27].

Considering described optimization framework for finding optimal symbol point locations for SNR-adaptive irregular constellation, it has become inevitable that the number of available constellation options is equal to the number of available MCSs. To illustrate the difference of the proposed MCS construction technique from the existing MCS framework, Fig. 2 gives a comparison over given three different SNR intervals for the conventional cases. In conventional cases, the same constellation can be used over more than one MCS whereas the proposed MCS design offers a wider range of optimized irregular

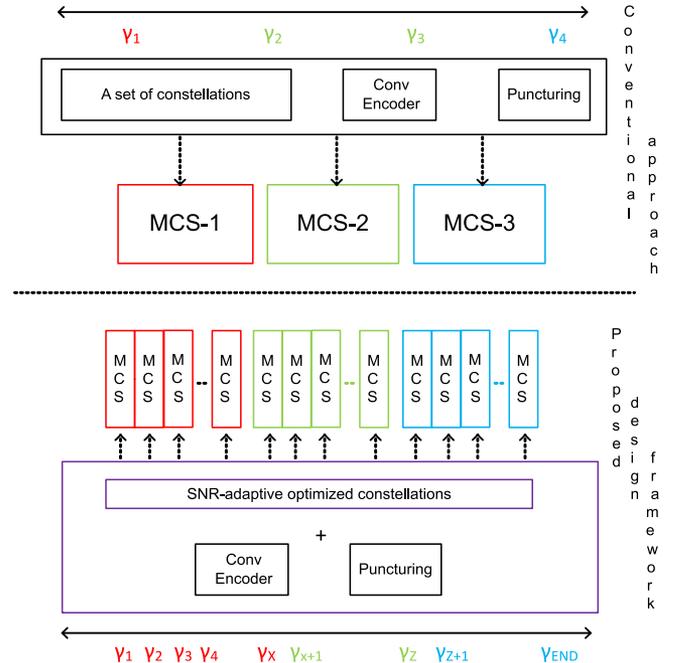

Fig. 2.   Basic principles of constructing conventional MCS (top) and the proposed SNR-adaptive MCS (bottom) techniques.

TABLE I
MCSs FOR MONTE CARLO SIMULATIONS

| MCS | Rate ($R$) | Modulation order ($M$) |
|-----|------------|------------------------|
| 1   | 1/2        | 16                     |
| 2   | 3/4        | 16                     |
| 3   | 3/4        | 64                     |

constellations which are specifically found for corresponding system parameter.

## IV.  SIMULATION RESULTS

Now, the performance of the proposed convolutionally coded SNR-adaptive transmission is investigated in terms of simulated BER and spectral efficiency (SE) values. In order to give a basic comparison between convolutionally coded SNR-adaptive and SNR-independent transmission models, a single-input single-output (SISO) system where rate-1/2 convolutional encoder $[5, 7]_8$ is employed is considered along with three different modulation and coding schemes (MCSs) given in Table I. In order to reach $3/4$ coding rate, the puncturing patterns $[1\ 1\ 0; 0\ 1\ 1]$ and $[1\ 1\ 0\ 1; 0\ 1\ 1\ 1]$ are used for MCS-2 and MCS-3, respectively [28].

### A.  Optimized Irregular Constellations

In order to motivate new approach for constructing MCSs along with convolutionally coded SNR-adaptive transmission model, the SEs are first plotted in Fig. 3 for $N_b = 920$, $m = 2$, $m = 4$, and

TABLE II
OPTIMIZED IRREGULAR CONSTELLATIONS $\chi(m, \overline{\gamma}[\text{dB}])$ FOR 16-ARY SIGNALING CASES FOR $\Omega = 1$

| $s_i \in \chi(m, \overline{\gamma}[\text{dB}])$ | MCS-1 $\chi$ (2, 12 dB) | MCS-1 $\chi$ (2, 18 dB) | MCS-2 $\chi$ (2, 12 dB) | MCS-2 $\chi$ (4, 18 dB) | MCS-1 & MCS-2 $\chi$-conventional |
|---|---|---|---|---|---|
| $s_0 - $ '0000' | 0.1358 + 0.6934i | 0.1344 + 0.7968i | 0.4180 + 0.2218i | 0.4011 − 0.2021i | 0.3162 + 0.3162i |
| $s_1 - $ '0001' | 0.2277 + 1.1093i | 0.1378 + 1.0616i | 0.3164 + 0.9437i | 0.4224 + 0.8205i | 0.3162 + 0.9487i |
| $s_2 - $ '0010' | 0.7812 + 0.2156i | 0.8691 + 0.1748i | 1.0867 + 0.2580i | 1.2082 − 0.1436i | 0.9487 + 0.3162i |
| $s_3 - $ '0011' | 1.1138 + 0.2908i | 1.0965 + 0.1874i | 0.7577 + 0.9934i | 0.8400 + 1.0048i | 0.9487 + 0.9487i |
| $s_4 - $ '0100' | 0.1536 − 0.6739i | 0.0952 − 0.7598i | 0.4231 − 0.1860i | 0.5018 + 0.1137i | 0.3162 − 0.3162i |
| $s_5 - $ '0101' | 0.2018 − 1.1258i | 0.1499 − 1.0601i | 0.3569 − 0.9393i | 0.5327 − 0.8708i | 0.3162 − 0.9487i |
| $s_6 - $ '0110' | 0.7925 − 0.2395i | 0.9095 − 0.1928i | 1.1119 − 0.2179i | 1.0799 + 0.0848i | 0.9487 − 0.3162i |
| $s_7 - $ '0111' | 1.1134 − 0.2804i | 1.0940 − 0.1970i | 0.8077 − 0.9658i | 0.8946 − 0.8777i | 0.9487 − 0.9487i |
| $s_8 - $ '1000' | −0.1485 + 0.7277i | −0.1192 + 0.7983i | −0.4002 + 0.1637i | −0.3726 − 0.1073i | −0.3162 + 0.3162i |
| $s_9 - $ '1001' | −0.2129 + 1.1209i | −0.1313 + 1.0539i | −0.3536 + 0.8912i | −0.4074 + 0.8057i | −0.3162 + 0.9487i |
| $s_{10} - $ '1010' | −0.7882 + 0.2383i | −0.8882 + 0.1854i | −0.1934 + 0.1829i | −1.1578 − 0.1654i | −0.9487 + 0.3162i |
| $s_{11} - $ '1011' | −1.1099 + 0.3083i | −1.0766 + 0.2004i | −0.8106 + 0.9501i | −0.8566 + 0.9038i | −0.9487 + 0.9487i |
| $s_{12} - $ '1100' | −0.1830 − 0.7536i | −0.1663 − 0.8178i | −0.3945 − 0.2104i | −0.4892 + 0.0895i | −0.3162 − 0.3162i |
| $s_{13} - $ '1101' | −0.1653 − 1.1270i | −0.139 − 1.0711i | −0.3249 − 0.9387i | −0.4336 − 0.8546i | −0.3162 − 0.9487i |
| $s_{14} - $ '1110' | −0.8090 − 0.2594i | −0.8722 − 0.2025i | −1.0691 − 0.2598i | −1.0589 + 0.1394i | −0.9487 − 0.3162i |
| $s_{15} - $ '1111' | −1.1028 − 0.2412i | −1.1185 − 0.1574i | −0.7736 − 0.9970i | −0.8248 − 0.9107i | −0.9487 − 0.9487i |

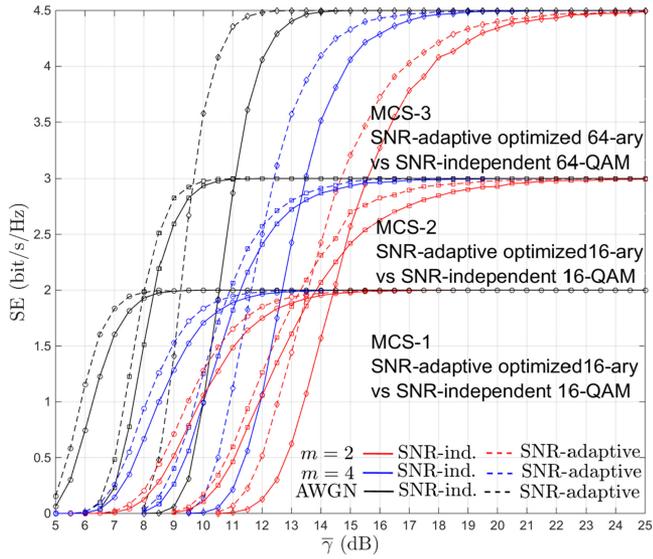

Fig. 3. Spectral efficiency comparison with different rates: SNR-adaptive optimized constellations vs. SNR-independent conventional $M$-QAM.

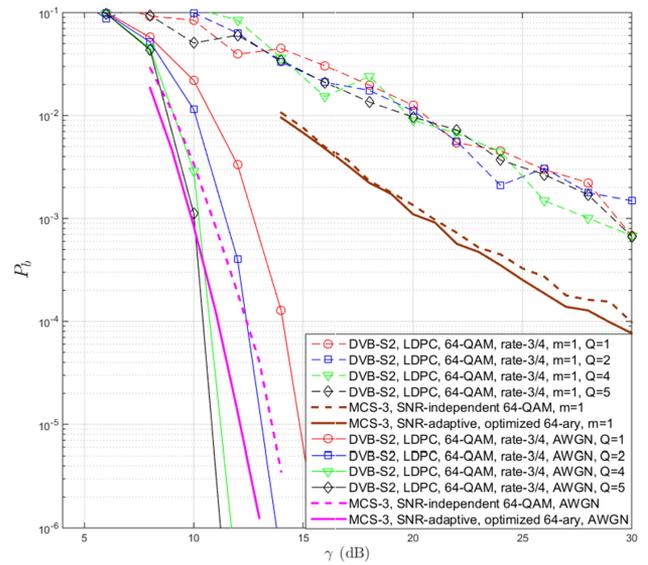

Fig. 4. Simulated BER comparison: SNR-adaptive optimized 64-ary constellation, SNR-independent conventional 64-QAM, and SNR-independent LDPC coded 64-QAM constellations for different decoding iterations ($Q$).

$m \to \infty$(AWGN) by using the formula [29], which is

$$\text{SE} = \log_2(M)(1 - (1 - P_b)^{N_b})R. \tag{8}$$

By choosing the outer curves for each pair of SE curves, considerable SE (spectral efficiency) increase can be obtained by using the proposed scheme. The variations in symbol point locations with respect to $m$ and $\overline{\gamma}$ can be easily seen in Table II.

Now, the performance evaluation of convolutionally coded SNR-adaptive transmission model is represented by comparing it with LDPC coded scenario where a LDPC encoder used in DVB-S2 and 802.16e standard is employed [30]. To sustain the original simulation platform in mentioned standard, $N_b$ is chosen as 19200 bits and 64-QAM constellation is used for SNR-independent cases over Rayleigh and AWGN channels. It can be seen from Fig. 4 that SNR-adaptive transmission model perform better than LDPC coded until four decoding iterations ($Q = 4$) over AWGN channels and considering its implementation complexity and decoding latency, SNR-adaptive transmission model can offer an alternative solution for low latency and low complexity communications. For Rayleigh cases, the superiority of LDPC coded transmission has been disappeared; then, fading parameter and average SNR-based convolutionally coded transmission outperform for each decoder iteration considered herein.

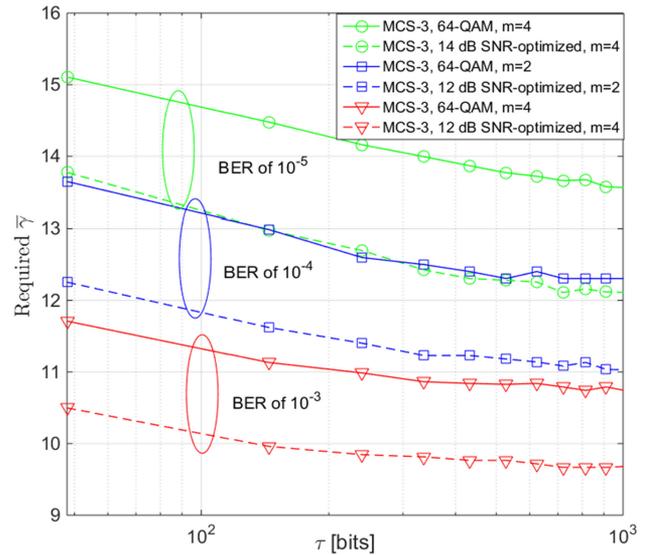

Fig. 5. The decoding latency comparison of convolutionally coded SNR-independent and SNR-adaptive transmission.

### B. Decoding Latency Improvement

Latency can be defined as the time interval from the moment the information sent from the transmitter to the completion of decoding process. Then, decoding latency can be obtained by excluding encoding time and channel delay from overall latency. In Fig. 5, the required $\bar{\gamma}$ values to obtain different threshold BER values for MCS-3 are plotted as a function of decoding latency in order to represent the advantage of the proposed SNR-adaptive transmission model in terms of decoding latency. For measuring the decoding latency, the window-length of back-search limit in Viterbi decoder, $\tau$, is selected for all mentioned cases [16]. The depicted curves show that the use of SNR-adaptive irregular constellation also brings a considerable advantage from decoding latency. It can be seen that the proposed SNR-adaptive framework can offer lower decoding delays under the same required $\bar{\gamma}$ to reach the same BER target value, which are at least several hundreds bits of lower decoding delays for given cases.

### V. CONCLUSION

Non-iterative/one shot-decoding characteristic and superiority performance under the certain decoding latency constraints motivate the use of convolutional encoders in some 5G use cases where low-latency communication and design complexity are the key criteria. Encouraged from this motivation, convolutionally coded SNR-adaptive transmission model is proposed and in this adaptive construction, the symbol locations vary with the received average SNR, channel characteristics, as well as the encoder properties. From this perspective, the proposed scheme allows working with different optimized constellations even for that selected coding rate and the modulation order stay the same, meanwhile it requires perfect knowledge of average link SNR and used constellation both transmitter and receiver. Under the assumption of perfect knowledge of them, it has been observed that considerable gains can be obtained with higher modulation order and spectral efficiency gains can be found in the order of 0.5–2.5 dB depending on the modulation level and channel characteristics.